# Optical Manipulation of Magnetic Vortex Visualized in situ by 4D Electron Microscopy

Xuewen Fu[1,*], Shawn D. Pollard[2], Bin Chen[3], Byung-Kuk Yoo[4], Hyunsoo Yang[2], Yimei Zhu[1,*]

**Affiliations:**

[1]Condensed Matter Physics and Material Science Department, Brookhaven National Laboratory, Upton, New York 11973, USA

[2]Department of Electrical and Computer Engineering, National University of Singapore, Singapore 117576, Singapore

[3]Center for Ultrafast Science and Technology, School of Chemistry and Chemical Engineering, Collaborative Innovation Center of IFSA, Shanghai Jiao Tong University, Shanghai 200240, China

[4]Physical Biology Center for Ultrafast Science and Technology, Arthur Amos Noyes Laboratory of Chemical Physics, California Institute of Technology, Pasadena, CA 91125, USA

*Corresponding author. E-mail: zhu@bnl.gov; xfu@bnl.gov

**Abstract:**

Understanding the fundamental dynamics of topological vortex and antivortex naturally formed in micro/nanoscale ferromagnetic building blocks under external perturbations is crucial to magnetic vortex based information processing and spintronic devices. All previous studies have focused on magnetic vortex-core switching via external magnetic fields, spin-polarized currents, or spin waves, which have largely prohibited the investigation of novel spin configurations that could emerge from the ground states in ferromagnetic disks and their underlying dynamics. Here, we report *in situ* visualization of femtosecond laser quenching induced magnetic vortex change in various symmetric ferromagnetic Permalloy disks by Lorentz phase imaging using 4D electron microscopy. Besides the switching of magnetic vortex chirality and polarity, we observed with distinct occurrence frequencies a plenitude of complex magnetic structures that have never been observed by magnetic field or current assisted switching. These complex magnetic structures consist of a number of newly created topological magnetic defects (vortex and antivortex) strictly conserving the topological winding number, demonstrating the direct impact of topological invariant on the magnetization dynamics in ferromagnetic disks. Their spin configurations show mirror or rotation symmetry due to the geometrical confinement of the disks. Combined micromagnetic simulations with the experimental observations reveal the underlying magnetization dynamics and formation mechanism of the optical quenching induced complex magnetic structures. Their distinct occurrence rates are pertinent to their formation-growth energetics and pinning effects at the disk edge. Based on these findings, we propose a paradigm of optical-quenching-assisted fast switching of vortex cores for the control of magnetic vortex based information recording and spintronic devices.

**Introduction**

Magnetic vortex[1,2] is one of the fundamental spin configurations occurring in thin micro/nanometer-sized ferromagnetic disk elements due to the confinement of spins imposed by geometrical restrictions[2,3]. It is one kind of topological magnetic defects characterized by two degrees of freedom[4]: (1) "chirality" ($c = \pm1$), the in-plane curling magnetization that can be clockwise or anticlockwise along the disk circumference; and (2) "polarity" ($p = \pm1$), the out-of-plane nanometer-sized core magnetization whose direction is either up or down. This topologically protected magnetic vortex cannot be continuously transferred into a defect-free state, and is therefore regarded as very robust quasiparticles against thermal fluctuation[5]. Such unique properties make magnetic vortex a promising candidate for high density, non-volatile magnetic memories and spintronic devices[6], because each magnetic vortex can store two bits of information by its chirality and polarity[7,8].

Fundamental understanding of the magnetization dynamics associated with the precise manipulation of the chirality and polarity of the magnetic vortex in ferromagnetic building blocks is important for its application in magnetic data-storage[7]. It is well known that the topological magnetic vortex in a ferromagnetic disk can be manipulated by external perturbations, such as pulsed magnetic fields[7,9,10], alternating magnetic fields[11,12], spin-polarized currents[7,13-15], and field-driven spin waves[16,17]. These external stimuli could drive the gyrotropic motion of the vortex core so that its polarization would be switched through the creation and subsequent annihilation of a magnetic vortex-antivortex pair[11,18-20]. Nevertheless, due to the gyrotropic motion of the vortex core prior to its switching, it is difficult to precisely determine when the core switching occurs, and thus limiting the ultimate investigation of the core switching dynamics that is important for designing vortex based data-storage devices. This gyrotropic motion also intrinsically restricts the speed of the magnetic vortex core switching.

Recently, it has been demonstrated the possibility of ultrafast magnetic switching in ferromagnetic thin films via photothermal-assisted femtosecond (fs) laser pulse excitation[21-25], where the single pulse rapidly heats up the ferromagnetic films close to their Curie temperatures and reduces the external magnetic field required for the magnetic reversal. Especially by using the inverse Faraday effect of circularly polarized fs-laser[26,27], it is even possible to realize all-optical magnetization switching in ferromagnetic films[28-32]. It has also been theoretically predicted the remarkable reduction of required magnetic switching field for a topological magnetic vortex core at a temperature closely below the Curie point[33] and the possibility of all-optical switching of magnetic vortex core[34] in ferromagnetic disks. This optically associated switching of magnetic vortex core does not involve the gyrotropic motion and thus possesses unique advantages in ultrafast magnetic recording. However, direct observation of magnetic vortex switching or change in geometrically confined ferromagnetic disks upon ultrafast laser quenching is rather challenging. Once achieved, it will provide a better fundamental understanding of the effects of topological feature, magnetization relaxation dynamics, and geometrical confinement on the magnetic vortex switching and its underlying mechanisms.

In this work, we report the *in situ* visualization of fs-laser quenching induced magnetic vortex configurations in symmetric ferromagnetic Permalloy (Py) disks by 4D electron microscopy (4D EM) operated in the Lorentz phase imaging mode. Besides the chirality and polarity switching of the magnetic vortex, a plenitude of complex metastable magnetic structures with distinct occurrence frequencies were observed in the Py disks with a fs-laser pulse excitation above a fluence of 10 mJ/cm$^2$. Different symmetric elements including circular, square and regularly triangular disks were designed for investigating the geometrical confinement effect. The observed magnetic structures consist of a number of newly created topological defects (vortex and antivortex) strictly restricted by the topological winding number, and their spin configurations show mirror or rotation symmetry due to the geometrical confinement of the disks. Micromagnetic simulations reproduce all the observed magnetic structures, revealing the underlying magnetization dynamics and the formation mechanisms. Based on the results, we propose a new paradigm of optical-quenching-assisted fast vortex core switching for the control of magnetic vortex based information processing and spintronic devices.

## Metastable magnetic structures induced by fs-laser pulse

The fs-laser pulse induced magnetic switching in three kinds of symmetric Py ($Ni_{81}Fe_{19}$) disks, including circle (diameter of 3 $\mu$m and 1.7 $\mu$m), square (edge length of 3 $\mu$m) and regular triangle (edge length of 1.7 $\mu$m), were studied (see Methods). To clearly resolve the magnetic vortex structures, we record the Fresnel images by using continuous electron beam of 4D EM in the Lorentz phase imaging mode (see Fig. 1a and Methods). Under out-of-focus condition of the Lorentz mode, the clockwise and counter-clockwise in-plane circling magnetizations of the magnetic vortices exert opposite Lorentz force on the imaging electrons, resulting in black and white contrasts of the vortex core[35-38], as schematically shown in Fig. 1b. Upon a fs-laser pulse excitation, the Py disk is rapidly heated up and subsequently followed by a fast cooling at a quenching rate of ca.$10^{12}$ K/s[39] passing through the thick substrate below (see the inset of Fig. 1b). Figure 1c present the typical Fresnel images of the magnetic structures in the circular, square and triangular Py disks before and after fs-laser pulse excitation with a fluence of 12 mJ/cm$^2$ (see also Movie S1 to S3). Based on the Lorentz contrast the corresponding spin configurations are schematically depicted in the right panel of each Fresnel image. After each fs-laser pulse excitation on the initial single magnetic vortex in all the three geometrical Py disks, the Lorentz contrast in their Fresnel images exhibit a high probability to reversal (see the first three columns in Fig. 1c), implying the repeatable switch of the chirality of the magnetic vortex by the rapid optical quenching. Below a threshold fluence of about 10 mJ/cm$^2$ a single fs-laser pulse excitation is insufficient to induce observable change of the initial magnetic vortex in the Py disks, since the pulse induced transient temperature of the Py disk is substantially lower than its Curie point (~ 850 K). Note that, the observed reversal of the magnetic vortex chirality behaves randomly in the experiment. Interestingly, besides the random reversal of the magnetic vortex chirality, we also observed, although less frequently, some complex magnetic structures consisting of several newly generated vortices, antivortices and domains in all the three geometries (see typical ones in the last column of Fig. 1c), which were never been observed by magnetic field or spin current assisted magnetic vortex switching. The magnetic antivortex is the topological counterpart of a magnetic vortex, which also contains a tiny core magnetized perpendicularly to the plane in the center and enclosed by two adjacent vortex structures[40,41]. It is discernible as a saddle point in the Fresnel image, namely, the cross of two Néel walls showing opposite Lorentz contrast (white and black)[39]. As in our work we did not intend to determine the polarity of the vortex cores, it is possible that the vortex polarity switching may also occur during the ultrafast optical quenching[25,34].

To understand the above interesting phenomenon, we repeated the same experiment on the three geometrical Py disks more than hundreds of times, in order to get statistically meaningful measurements of the fs-laser pulse induced metastable magnetic structures so that the underlying mechanism could be retrieved. The typical Fresnel images of the observed metastable magnetic structures are displayed in the middle panel of each subpanel in Fig. 2, while their corresponding spin configurations and occurrence frequency distributions are shown in the panels below and above, respectively. Each Fresnel image was acquired promptly after each fs-laser pulse excitation (fluence of 12 mJ/cm$^2$). In each statistic histogram, the most frequently observed single clockwise and counter-clockwise vortex structures with opposite Lorentz contrast were counted separately, while the other complex magnetic structures with opposite Lorentz contrast but having the same spin configuration were added together. In all the three disks the counter-clockwise and clockwise single vortex states occur randomly with the similar occurrence frequency, which is overwhelmingly higher than that of other complex magnetic structures. This nearly 90% occurrence frequency of the single vortex state (including both clockwise and counter-clockwise ones) also verifies the lowest energy and highest stability of it in the symmetric Py disks. All the other complex magnetic structures consist of a number of newly generated vortices, antivortices, domains and pairs of Néel walls, where the antivortex is located between two vortices with the same chirality. Intriguingly, the spatial distribution of their spin configurations exhibits striking mirror or rotation symmetry (see Fig. 2), which is probably due to the confinement of the geometrical symmetry of the Py disks. For a magnetic structure containing a number of topological defects in a symmetric ferromagnetic disk, the symmetric spin configuration will aid to reduce the total energy of the system. Note that, these complex metastable magnetic structures are much more difficult to form in smaller circular Py disks (1.7 µm diameter) under the same quenching condition (see Movie S4), where only single magnetic vortex structure forms, implying the dimensionality affects the magnetization dynamics and final magnetic states.

To unravel the origin of the different occurrences of the complex metastable magnetic structures, we further carried out the fs-laser quenching experiments with a higher laser fluence of 16 mJ/cm$^2$. At this fluence, other more complex, symmetric metastable magnetic structures consisting of a larger number of topological defects were observed in all the three geometries (see Fig. S1 to S3). For both the circular and square disks, the magnetic structures comprise various vortices up to 6 were observed, while for the triangular one the most complex magnetic structure only contains 4 vortices, which is probably due to its lower geometrical symmetry. Note that, most of these complex magnetic configurations have never been observed in ferromagnetic disks with other external stimuli, such as annealing, magnetic field, spin-polarized current and spin wave. Basically, the more complex magnetic configurations (with more topological defects) show lower occurrence frequency in the fs-laser quenching experiment (Fig. S1 to S3). The occurrence of additional more complex, symmetric magnetic structures at this high fluence is mainly due to the strong laser heating induced crystallite change in the Py disk, which will be discussed later.

The magnetic vortex and antivortex are both topological defects with the local magnetization rotating by 360° on a closed loop around the tiny core, which can be characterized by a topological winding number $w = 1/2\pi \oint \nabla\alpha \cdot dS$ ($w = \pm 1$ for vortex and antivortex, respectively), where $\alpha$ is the local orientation of the magnetization vector, $S$ is an arbitrary integral loop containing the tiny core[18]. The topological winding number has been theoretically predicted to have a direct impact on the magnetization dynamics[40,42]. Because of the spatial symmetry breaking at the edge of the Py disks, each cross of a pair of Néel walls with opposite Lorentz contrast at the edge can be considered as a half-antivortex and its topological winding number $w$ turns out to be -1/2. One would find that the sum of the total topological winding numbers for each observed metastable magnetic structure in all the three geometrical disks is equal to 1, which is the same as that of their initial single magnetic vortex state. Namely, the generation of new vortices and antivortices during the fs-laser quenching conserves the topological winding number of the Py disks, i.e., is strictly restricted by the topological invariant. This topological feature is similar to that of the light induced magnetic network in homogeneous ferromagnetic iron thin films, where the vortex-antivortex generates in pairs and follows the universal behavior within the framework of Kibble-Zurek mechanism[39].

**Magnetization dynamics and formation mechanism**

To understand the formation mechanism and topological feature of the symmetric magnetic structures induced by fs-laser pulse quenching, we performed finite-element micromagnetic simulations on these three geometrical Py disks based on the Landau-Lifshitz-Gilbert equation with Langevin dynamics[29] to reveal the underlying magnetization dynamics (see Methods). For the micromagnetic simulations we consider following scenarios: (1) the fs-laser pulse only interacts with the magnetization via the photothermal effect; (2) the fs-pulse heats the Py disk above the Curie point and randomizes the local magnetization, namely, melts the electronic spin structures, but without altering the integrity of the lattice; and (3) each fs-laser pulse excitation may result in different random magnetization seeds in the melted spin system. Under these conditions the optical quenching induced magnetization dynamics in the Py disks can be understood as follows. Upon a fs-laser pulse excitation above the threshold fluence for the Curie temperature, the thermal energy of the electronic system in the Py disk is rapidly increased, creating a thermal bath for the spin system. This sharp increase in thermal energy of the spin system leads to a rapid and full demagnetization, namely, spin melt of the initial magnetization in the Py disk within several picoseconds[38]. The subsequent energy transfer from the spin system to the lattice via electron-phonon coupling[29] leads to a rapid decrease of the temperature with a cooling rate up to $10^{12}$ K/s[39] to below Curie point initiating the remagnetization process of the spin system.

We used the exact dimensions of the samples in the micromagnetic simulations, and more than 25 runs of the numerical simulation were performed on each Py disk. The bottom panel of each subfigure in Fig. 3 presents the typical spin configurations in each Py disk generated by the micromagnetic simulation, and their corresponding occurrence frequencies (plotted in pink bars) and energies are plotted together in the top panel. For comparison, the corresponding occurrence frequency distributions of the magnetic structures measured by the experiments are also plotted in blue bars. Here the single magnetic vortex states with opposite chirality were counted together. Clearly, all the magnetic structures observed in the experiment (laser fluence of 12 mJ/cm$^2$) are well reproduced by the micromagnetic simulations for all the three geometrical disks, and their occurrence frequencies agree as well (Fig. 3).

In all three geometrical disks, the single magnetic vortex state always exhibits the highest occurrence due to its lowest energy (~ $6.98 \times 10^{-17}$ J for circle, ~ $7.32 \times 10^{-17}$ J for square, and ~ $9.52 \times 10^{-17}$ J for triangle). While the occurrence of other complex magnetic structures generally decreases with their energy (see Fig. 3). Specifically, for the magnetic structures in circular disk, their energy nearly increases linearly with increasing the number of the contained topological magnetic defects (vortex) and their occurrence frequency decreases monotonously (Fig. 3a). While for both the square and triangular disks, the energy increase levels off for the magnetic structures with more than three topological magnetic defects (vortex); and counterintuitively, the magnetic structure containing two vortices with lower energy even exhibits smaller occurrence frequencies than that of the magnetic structure containing three vortices with higher energy (Fig. 3b and 3c). This abnormal behavior is probably due to the much smaller energy barrier for the magnetic structure containing two vortices to overcome during the magnetization relaxation after the optical quenching. Based on the micromagnetic simulation, the relative energy barriers for different magnetic structures during the magnetization relaxation in a square disk are schematically depicted in Fig. 4. Because of the much smaller energy barrier, the metastable magnetic structure containing two vortices strongly prefers to relax to the single magnetic vortex state, resulting in the lower occurrence than the magnetic structure with three vortices in both square and triangular disks (Fig. 3b and 3c).

To unravel the factors that determine the final magnetic structures, the initial remagnetization process of the random magnetization seeds, especially the role of the spin pinning at the edge defects of the Py disks, was considered in our micromagnetic simulation. In a perfectly circular disk with smooth edge, the single magnetic vortex state is strongly preferred after a fs-laser pulse excitation. Due to the confinement of the geometrical symmetry of the disk and the spin pinning at the edge defects (or edge roughness), the finally magnetic structures prefer to form symmetric configuration to reduce the system total energy. For further discussion, three exemplary time-dependent magnetization evolutions for the formation of the single magnetic vortex state, the metastable magnetic structures with two and three vortices in a triangular disk are respectively presented in Fig. 5 (see also Movie S5 to S7). After a fs-laser pulse induced randomization of the initial magnetization in the disk, the melted spin system starts to remagnetize when the temperature cools down to below the Curie point in several picoseconds and a number of vortices, antivortices, domains and half antivortices at the disk edge are formed. Note that, these initially formed topological magnetic defects converse the topological winding number ($w = 1$) of the initial single magnetic vortex sate in the Py disk. With time elapse, the adjacent vortices and antivortices inside the disk move spirally and approach each other, and then annihilate in pairs; while for the half antivortices at the disk edge, a nearby inside vortex moves towards the center of two adjacent half antivortices and they annihilate once meet together, as indicated by the colored circles in Fig. 5. Depending on the relative core polarization of the adjacent vortex-antivortex pair, their annihilation process is either a continuous transformation of the magnetization (parallel), or involving nucleation and propagation of a Bloch point causing a burstlike emission of spin waves (antiparallel)[18]. During this incessant vortex-antivortex pair annihilation process, the energy of the spin system continuously decreases, until the system relaxes to the single magnetic vortex state with lowest energy (see Fig. 5a).

Because of the spin pinning at the disk edge (indicated by the blue and pink arrows in Fig. 5b and 5c; see also Fig. S4) the half-antivortices at these pinning sites cannot move freely, thus the two adjacent half antivortices at these pinning sites cannot annihilate with a nearby vortex. In such case, the spin system relaxes to a symmetric multivortex state due to the confinement of the symmetric geometry of the disk (Fig. 5b and 5c). The magnetization dynamics and the formation mechanisms of other metastable complex magnetic structures, including those in the circular and square Py disks, have the similar features (see Movie S8 to S13). The strong spin pinning effect at the disk edge could also interpret the observation of additional more complex, symmetric magnetic structures in our experiment at a higher laser fluence of 16 mJ/cm$^2$ (see Fig. S1 to S3). At this high fluence, the strong heating effect would cause lattice grain growth within the Py disk (see Fig. S5), which would induce larger grain boundaries, i.e., roughness at the disk edge, and thus more strong spin pinning sites, resulting in more complex magnetic structures (Fig. S1 to S3).

## Optical-quenching-assisted magnetic vortex switching for information recording

Thermally assisted magnetization reversal has been proposed as one of the most promising way to enable high density magnetic recording[24,43], where the large anisotropy values required for the stability of the recording media films are transiently reduced by laser heating and the required magnetic switching field is markedly declined. Recently, Lebecki and Nowak[33] have theoretically studied the temperature impact on the magnetic switching field of a magnetic vortex core in a circular ferromagnetic disk based on the Landau-Lifshitz-Bloch equation[44] incorporating with the thermal effect. As shown in their prediction, the orthogonal magnetic field required for vortex core switching dramatically decreases with increasing temperature due to the lower energy barrier at higher temperature, which may even vanish at a temperature slightly below the Curie point (see Fig. 3 and 5 in Ref. 33). In contrast, our experimental results clearly demonstrate that both the magnetic vortex switching and the final magnetic state are random after a fs-laser pulse excitation (see Fig. 2), implying the purely photothermal switching of a magnetic vortex is possible but uncontrollable, which will seriously hinder the practical application. Nevertheless, a fs-laser pulse with a proper fluence could initiate a sharp increase of temperature in the Py disk above its Curie point to induce a transient non-equilibrium paramagnetic state (spin melted state). During this transient paramagnetic period, if a small external orthogonal magnetic field is applied one could easily control the spin direction of the system and initiate the polarization of the magnetic vortex that formed in the subsequent magnetization relaxation process due to the ultrafast quenching.

Based on our experimental observations together with the simulation results, we propose a new paradigm of optical-quenching-assisted fast switching of the magnetic vortex polarity for the control of magnetic vortex based information recording and spintronic devices, as schematically shown in Fig. 6. A linear polarized fs-laser laser pulse to transiently demagnetize the initial magnetic vortex in the Py disk, and a small orthogonal magnetic field pulse (with duration above tens of picoseconds) to set the polarization of the newly formed magnetic vortex are simultaneously applied. In principle, the strength of this external magnetic field pulse should be much smaller than that (500 mT) of the orthogonal magnetic field required for the conventional quasistatic switching of a magnetic vortex core[45,46], which could be handily determined by magnetic force microscope measurements. It should be mentioned that, for this proposed optical-quenching-assisted magnetic vortex based information recording paradigm, the following factors need to be considered carefully: (1) the fluence of the fs-laser pulse is crucial, which should be able to instantaneously drive the system above the spin transition temperature, but below the temperature that would cause the crystallite damage in the Py disk. Too strong fs-laser pulse excitation would induce lattice change and cause the growth of crystallites in the Py disk (see Fig. S5), resulting in additional spin pinning sites that would frustrate the magnetization relaxation. (2) The dimension and the edge smoothness of the Py disk are also very important. As shown in our results, the small circular Py disks (1.7 µm diameter) strongly prefer to single magnetic vortex state after each fs-laser pulse quenching (see Movie S4). The smooth disk edge would reduce the spin pinning effects during the magnetization relaxation. Therefore, proper design of nanoscale Py disk with smooth edge and uniform small crystallites would be greatly helpful to improve the stability and reliability of the optical-quenching-assisted magnetic vortex switching, which would offer the new possibility for magnetic vortex based high density information recording with fast writing rates, but consuming much less power.

In conclusion, by using the unique ultrafast quenching rate of up to $10^{12}$ K/s of a fs-laser pulse excitation, we observed a plenitude of new metastable magnetic structures in three types of symmetric, micrometer-sized Py disks by 4D EM operated in the Lorentz phase imaging mode. These metastable magnetic structures consist of a number of newly created topological magnetic defects strictly restricted by the topological invariants, which were not observed previously in magnetic field or spin-current assisted magnetic vortex switching. Due to the confinement of the disk geometrical symmetry, their spin configurations show apparent mirror or rotation symmetry. Micromagnetic simulations revealed the underlying magnetization dynamics of all the observed magnetic structures, and the dependence of their occurrence frequencies on their energetics and pinning effects at the disk edge. Our results provide new insights into the fundamental spin switching dynamics in symmetric Py disks under fs-laser pulse quenching, which offers a guidance for the design of optical-quenching-assisted fast switching of topological vortices for vortex based information recording and spintronic devices.

## Methods

**Preparation of Py disks**

The samples studied in our experiments were prepared by electron beam evaporation of a layer of Py ($Ni_{81}Fe_{19}$) film (30 nm thickness, ca. 10 nm grain size) onto silicon nitride membrane (100 nm thickness, 300 $\mu$m × 300 $\mu$m window area) on silicon frame with prior prepared symmetric patterns using photolithography and liftoff process. Three kinds of symmetric disks were prepared: circle (diameter of 3 $\mu$m and 1.7 $\mu$m), square (edge length of 3 $\mu$m) and regular triangle (edge length of 1.7 $\mu$m). The corners of the square and regular triangle disks are intentionally made of an arc shape to avoid artificial singularity in spin switching (see Fig. 1c).

**Lorentz phase imaging of fs-laser pulse induced magnetic structures**

To image the fs-laser quenching induced changes of magnetic structure in the Py disks, we performed out-of-focus Fresnel phase imaging in a 4D EM[47-50] operated at Lorentz-mode condition[51,52]. To obtain high Lorentz contrast, the images were collected by using the continuous electron beam of the 4D EM rather than pulsed electrons. Linearly polarized green fs-laser pulses (520 nm, 40 $\mu$m focal spot size, 350 fs pulse duration) were used for excitation, which were generated from infrared fs-laser pulses (1040 nm, 350 fs pulse duration) by a second harmonic generation. The in-plane circular magnetizations (clockwise or counter-clockwise) of the magnetic vortices exert opposite Lorentz force on the imaging electrons, resulting in contrasts, or phase shift of the electron beam, related to the vortex core[36-38], respectively. The high throughput Fresnel phase imaging allows to investigate the fs-laser quenching induced magnetization changes at nanometer scale and the statistical properties of the resulting magnetic structures.

**Micromagnetic simulation**

Micromagnetic simulations were performed using micromagnetic software (OOMMF).[53] A saturation magnetization, $M_s$, of 800 kA/m, exchange stiffness, $A$, of 13 pJ/m, and a Gilbert damping, α, of 0.01, consistent with typical values of Py, were used. A stopping condition of $dM/dt = 0.1$ was used to ensure convergence. The cell size was set to 5 nm × 5 nm × 30 nm. The sample structure for each geometry was determined by creating a binary mask from TEM images for each. To determine possible metastable domain configurations, 25 runs for each geometry were performed using different initial random magnetic configurations, and then allowed to relax. Following relaxation, the energies and static domain configurations were recorded.

## Acknowledgements

We acknowledge Caltech for providing the access to 4D electron microscopy facility for this study. This work is supported by the Materials Science and Engineering Divisions, Office of Basic Energy Sciences of the U.S. Department of Energy under Contract No. DESC0012704. We wish to thank Dr. J. S. Baskin for very helpful discussion and help on the Lorentz phase electron microscopy measurement with *in situ* fs-laser excitation. We also wish to thank J. A. Garlow for fruitful discussion on the Lorentz phase imaging measurement.

## Author contributions:

Y. Z. and X. F. conceived the research project. X. F., B. C. and B. K. Y. did the experimental measurements. S. D. P. prepared the samples. X. F. and S. D. P. did the data analysis with input from Y. Z. S. D. P. developed the model and performed the numerical simulations. All the authors contributed to the discussion and the writing of the manuscript.

## Additional information

Supplementary information is available in the online version of the paper, including Fig. S1 to S5 and Movie S1 to S13.

## Competing financial interests

The authors declare no competing financial interests.

**Figure legends**

**Fig. 1. fs-laser pulse quenching of magnetic vortex in Py disks. a**, Sketch of imaging the fs-laser pulse induced change of spin configuration in a ferromagnetic Py disk by 4D EM operated in Lorentz phase mode with continues electron beam. The green fs-laser pulse (520 nm, 350 fs pulse duration) is focused to 40 $\mu$m on the sample. **b**, Schematic Lorentz contrast reverse mechanism of a magnetic vortex in a circular Py disk before and after a fs-laser pulse excitation due to the change of spin chirality. Because of the opposite Lorentz force of the imaging electrons impinging on the sample, the Lorentz contrast of a vortex core can be either black or white. The inset depicts the typical transient temperature evolutions after a fs-laser excitation (calculated by a two-temperature model[39]) in both the Py disk and the silicon nitride substrate ($T_C$ is the Curie point of Py disk, $T_R$ is the room temperature, laser fluence of 12 mJ/cm$^2$). **c**, fs-laser pulse induced variation of magnetic vortex in circular, square and regularly triangular Py disks. The right panel of each Fresnel image schematically depicts the corresponding spin configuration. The blue and red dashed lines correspond to the white and black Lorentz contrasts, respectively. While the blue and red dots to the counter-clockwise and clockwise vortices, individually. The green dots mark the magnetic antivortex. The same notes are used in all the following figures.

**Fig. 2. Occurrence frequency distribution of the fs-laser pulse induced magnetic structures in three geometrical Py disks. a-c**, Frequency distribution of the fs-laser pulse (fluence of 12 mJ/cm$^2$) induced spin configurations in circular, square and triangular Py disks, respectively. The bottom panel in each subfigure shows the typical Fresnel images of the experiments, and the middle panel schematically shows their corresponding spin configurations. The most frequent single clockwise and counter-clockwise vortex structures with opposite Lorentz imaging contrast were counted separately, while the other magnetic structures with opposite Lorentz imaging contrast but having the same spin configuration were added together. The inset in each subfigure denotes the fs-laser pulse quenching process in the Py disks.

**Fig. 3. Comparison of micromagnetic simulations with experimental observations. a-c**, Simulation results of the occurrence frequency distribution and energies of the fs-laser pulse (fluence of 12 mJ/cm$^2$) induced magnetic structures in circular, square and triangular Py disks, respectively. The bottom panel in each subfigure shows the typical results of possible magnetic structures obtained by the micromagnetic simulations (pink bars). The corresponding experiment determined occurrence frequency distribution of the fs-laser pulse induced magnetic structures are also plotted in blue bars for comparison. The simulation results reproduce well the experimental results, except one magnetic structure in the triangle disk (indicated by the dashed red circle in Fig. 3c).

**Fig. 4. Schematic of relative energy barriers for different metastable magnetic structures to overcome during the magnetization relaxation in a square Py disk. The magnetic structure containing two vortices has a smaller barrier to overcome comparing with other metastable magnetic states.**

**Fig. 5. Typical magnetization dynamics in Py disk after a fs-laser pulse quenching**. Snapshots of the magnetization dynamics during the formation of different magnetic structures in a triangular Py disk at different times after a fs-laser pulse excitation: **a**, formation a single magnetic vortex state; **b**, formation of a magnetic structure with two vortices; **c**, formation of a magnetic structure with three vortices. The laser fluence is 12 mJ/cm$^2$. The vortices and antivortices (including the half-antivortices indicated by the green half dots at the disk edge) in the different colored circles indicate the magnetic vortex-antivortex pairs that annihilate during the magnetization relaxation process. The blue and pink arrows indicate the spin pining sites at the disk edge.

**Fig. 6. A paradigm of optical-quenching-assisted magnetic vortex based information recording process.** Left: Schematic of the optical-quenching-assisted magnetic vortex based information recording system, where a linear polarized fs-laser laser pulse is used to transiently demagnetize the initial

magnetic vortex and another synchronized orthogonal small magnetic field pulse is used to set the polarization of the newly formed magnetic vortex. Right: sketch for the working mechanism of the optical-assisted magnetic vortex based information recording process. The data information "1" and "0" are recorded by the polarity (up and down) of the magnetic vortex. The fluence of the fs-laser pulse should be controlled above the threshold for spin melting, but below that for causing change of crystallites in the ferromagnetic disk.

# Figures

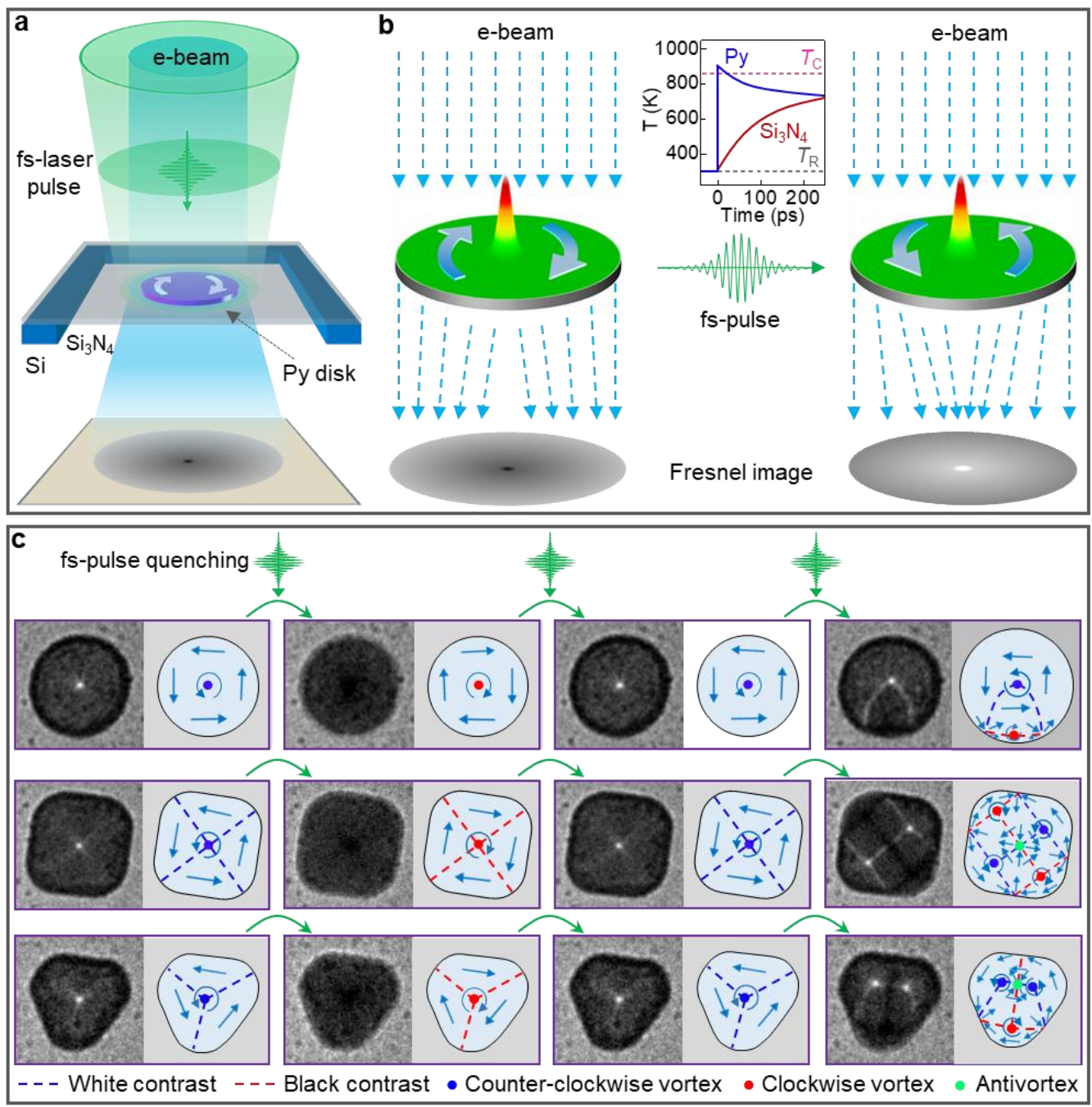

**Figure 1**

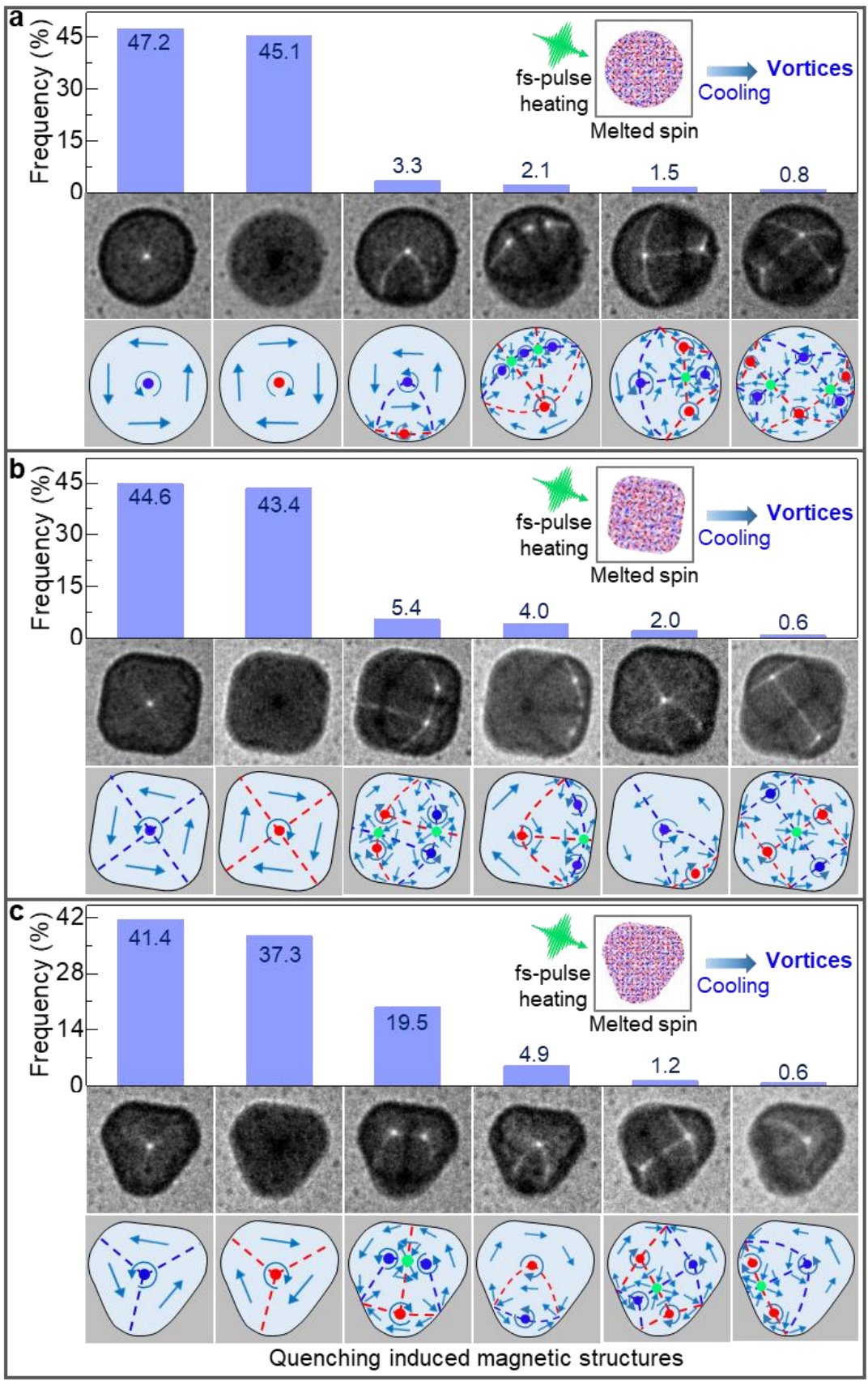

Quenching induced magnetic structures

**Figure 2**

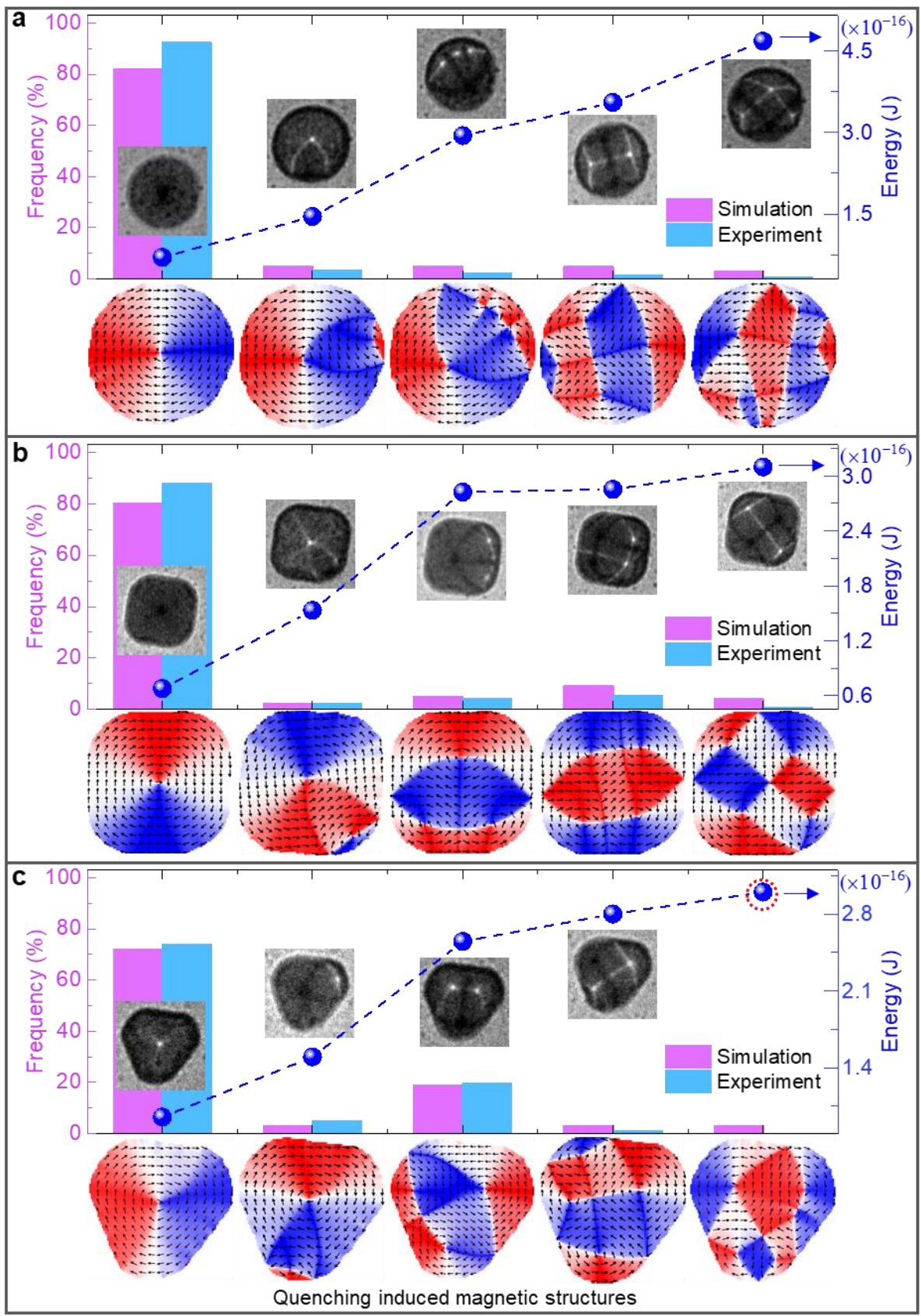

**Figure 3**

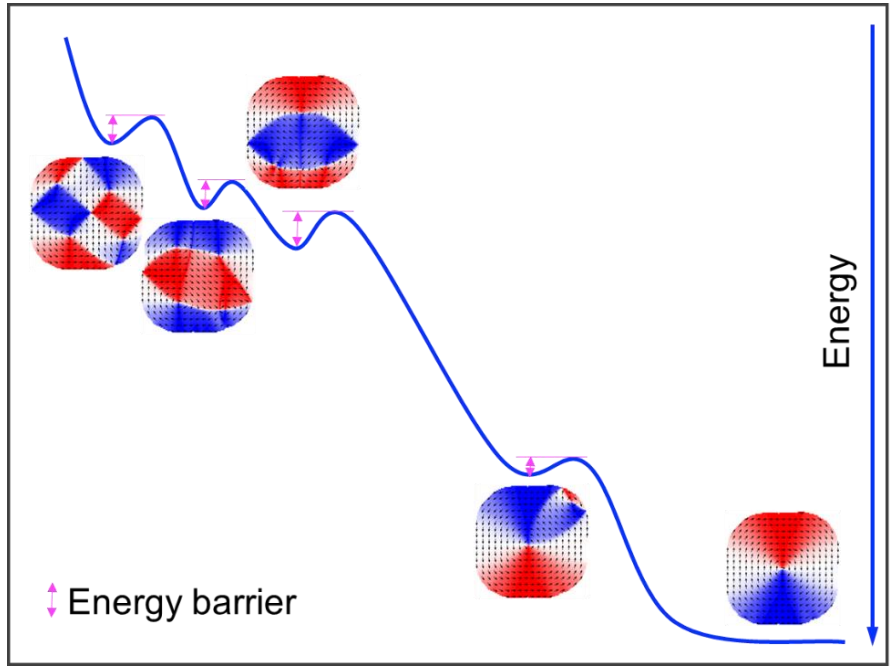

**Figure 4**

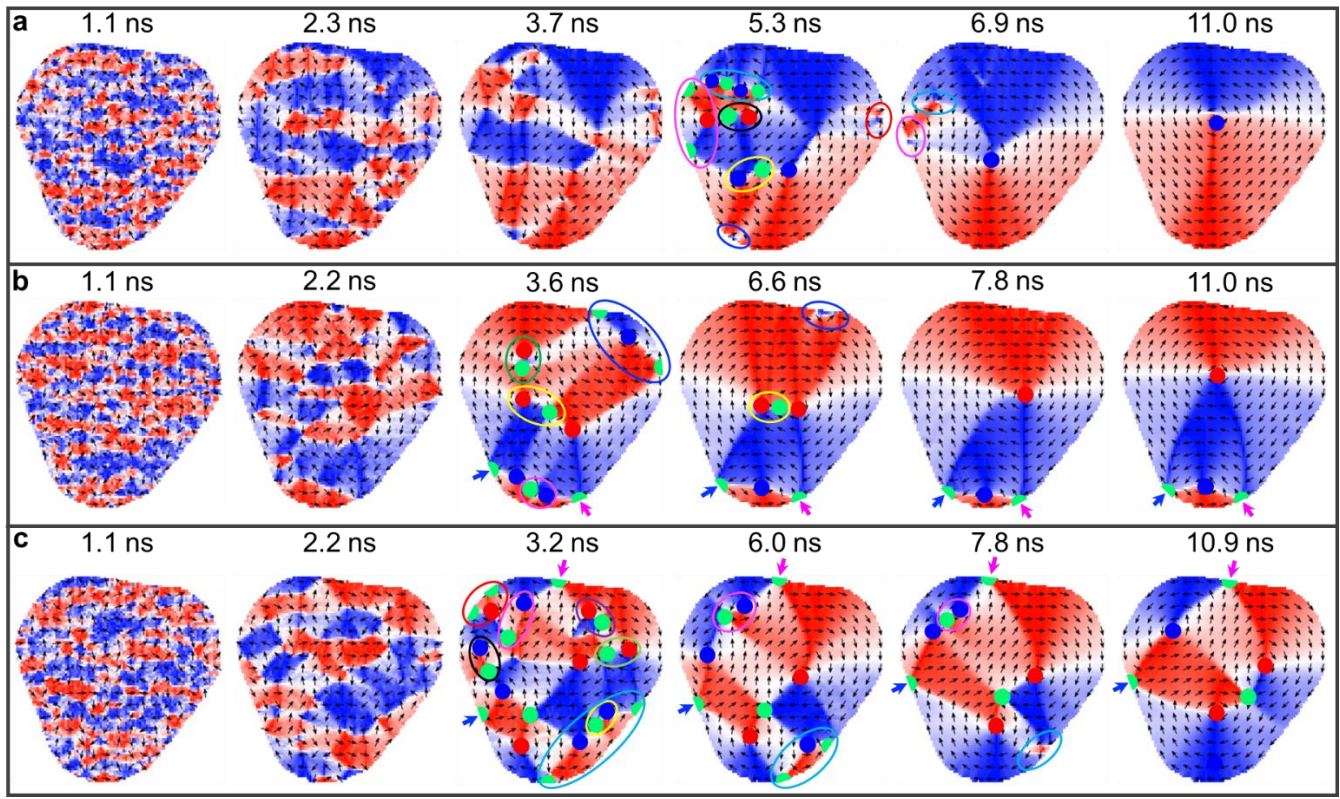

**Figure 5**

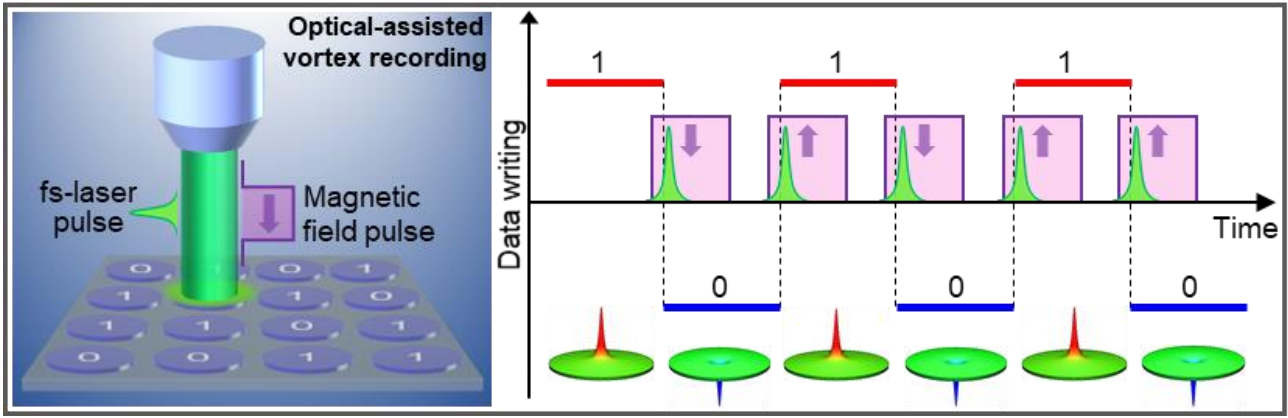

**Figure 6**

# *Supplementary Information for*

## Optical Manipulation of Magnetic Vortex Visualized in situ by 4D Electron Microscopy

Xuewen Fu[*], Shawn D. Pollard, Bin Chen, Byung-Kuk Yoo, Hyunsoo Yang, Yimei Zhu[*]

*Corresponding author. E-mail: zhu@bnl.gov; xfu@bnl.gov

**This PDF file includes:**

Figs. S1 to S5

Captions for Movies S1 to S13

**Other Supplementary Information for this manuscript includes the following:**

Movies S1 to S13

**Figs. S1 to S5**

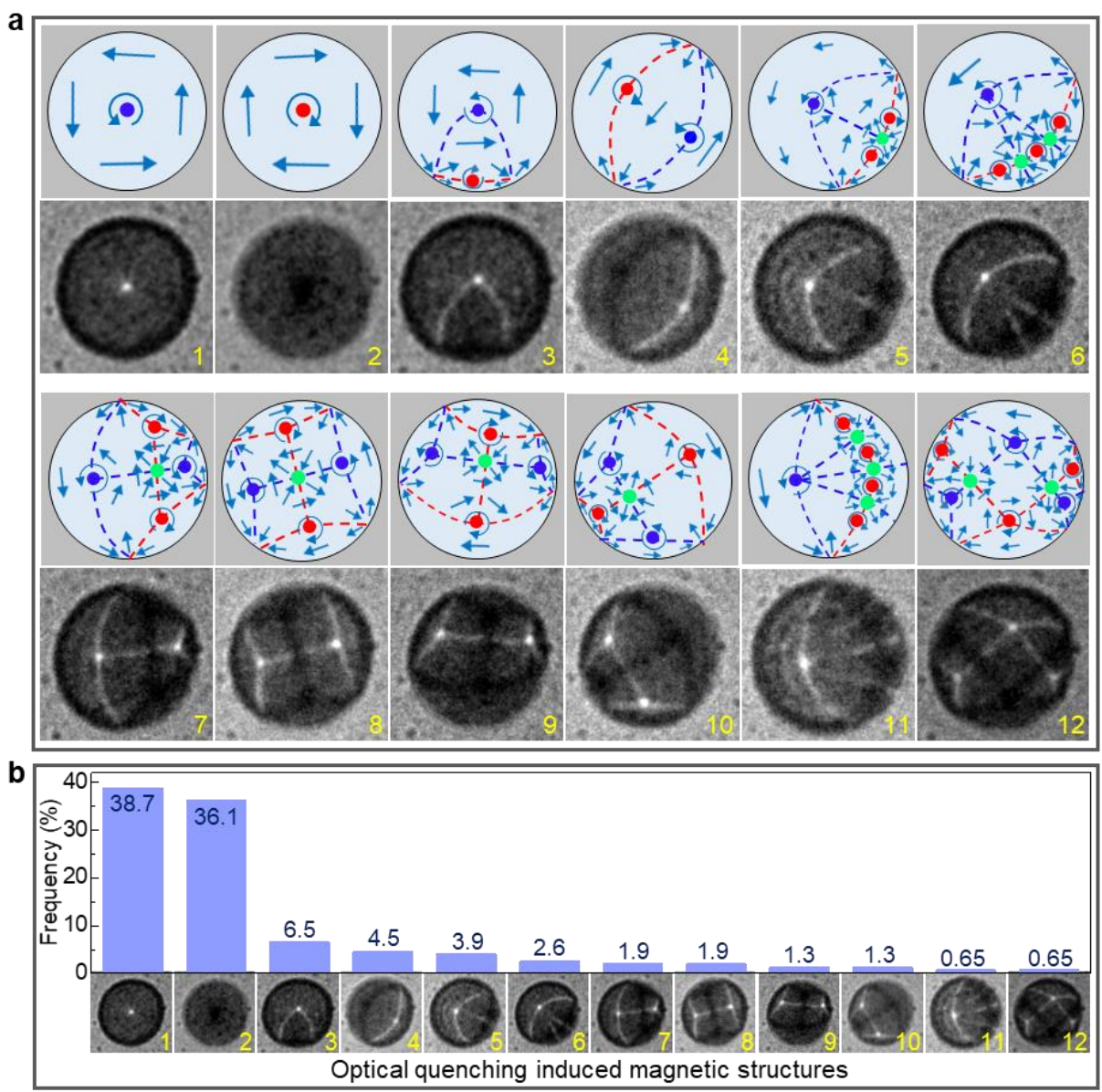

**Fig. S1. Occurrence frequency distribution of fs-laser pulse induced magnetic structures in a circular Py disk at a fluence of 16 mJ/cm$^2$.** (**a**) Bottom panel: Fresnel images of the observed magnetic structures in the circular Py disk (diameter of 3 $\mu$m); Top panel: corresponding schematic magnetization configurations. (**b**) Occurrence frequency distribution of the fs-laser pulse induced different magnetic structures.

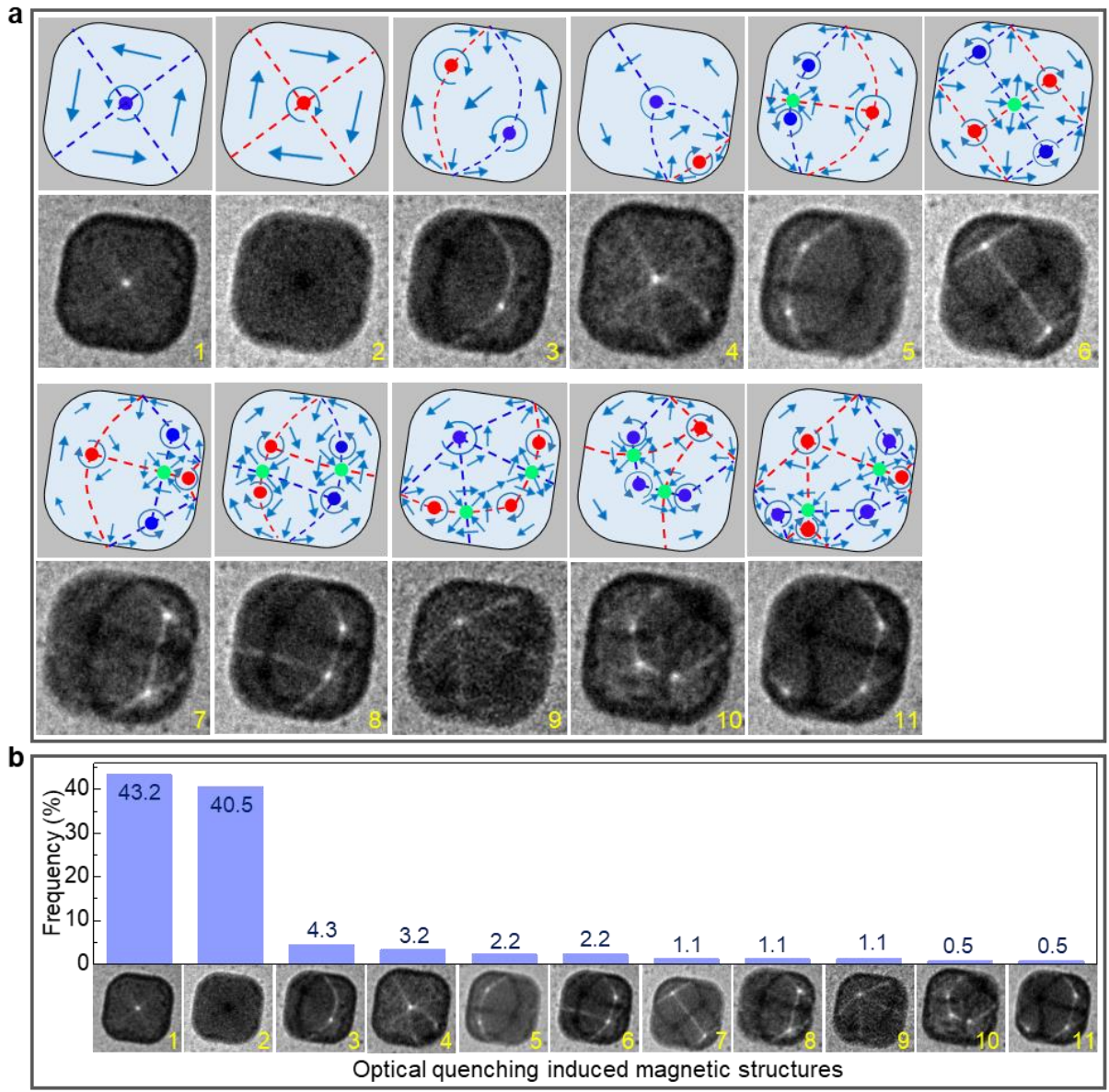

**Fig. S2. Occurrence frequency distribution of fs-laser pulse induced magnetic structures in a square Py disk at a fluence of 16 mJ/cm$^2$.** (**a**) Bottom panel: Fresnel images of the observed magnetic structures in the square Py disk (edge length of 3 $\mu$m); Top panel: corresponding schematic magnetization configurations. (**b**) Occurrence frequency distribution of the fs-laser pulse induced different magnetic structures.

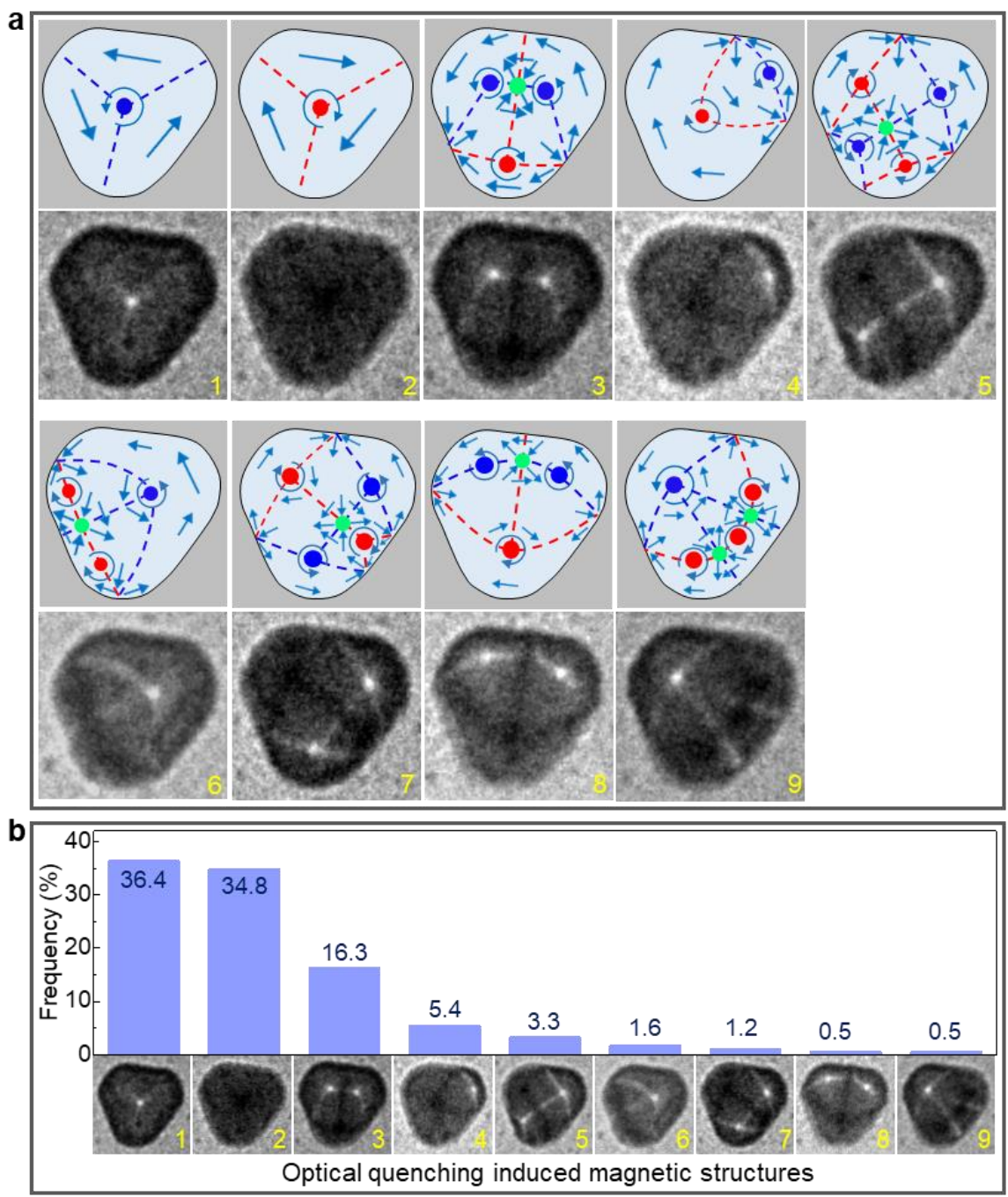

**Fig. S3. Occurrence frequency distribution of fs-laser pulse induced magnetic structures in a triangular Py disk at a fluence of 16 mJ/cm$^2$.** (**a**) Bottom panel: Fresnel images of the observed magnetic structures in the triangular Py disk (edge length of 1.7 $\mu$m); Top panel: corresponding schematic magnetization configurations. (**b**) Occurrence frequency distribution of the fs-laser pulse induced different magnetic structures.

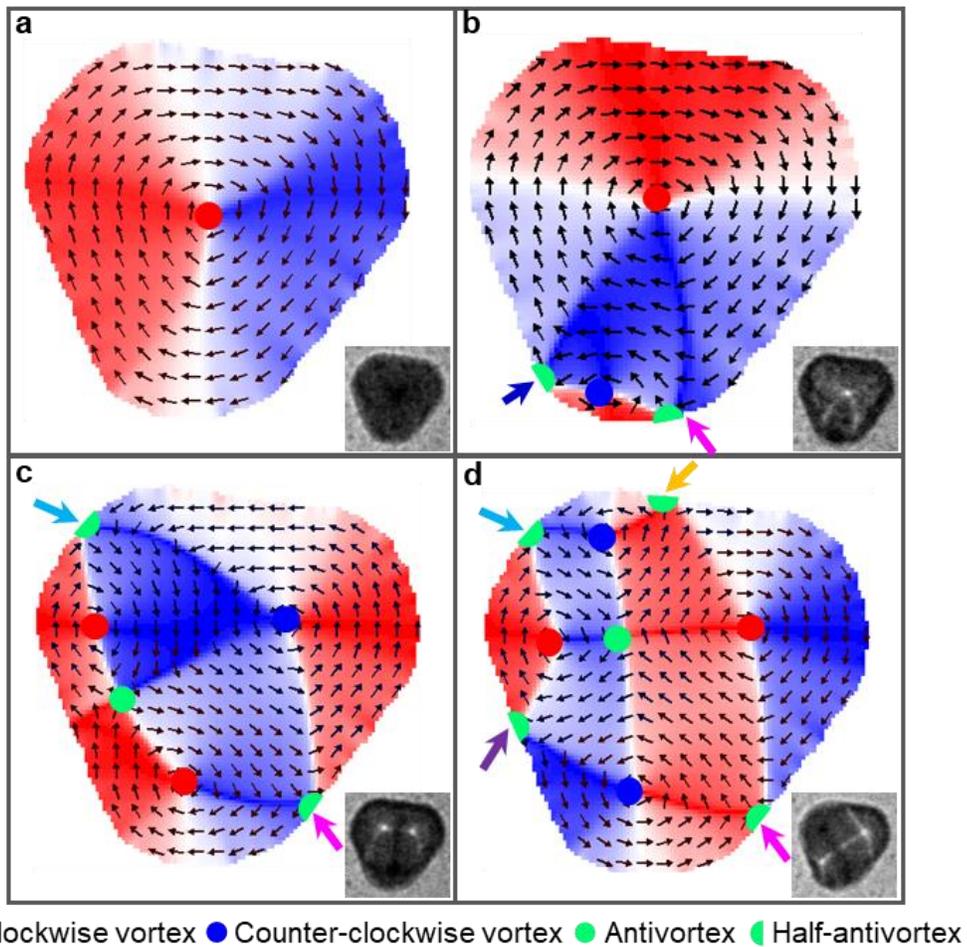

**Fig. S4. Typical magnetic structures in triangular Py disks (edge length of 1.7 $\mu$m) determined by micromagnetic simulation to show the pinning sites at the disk edge. The colored arrows at the disk edge show the pinning sites. The insets show their corresponding Fresnel images.**

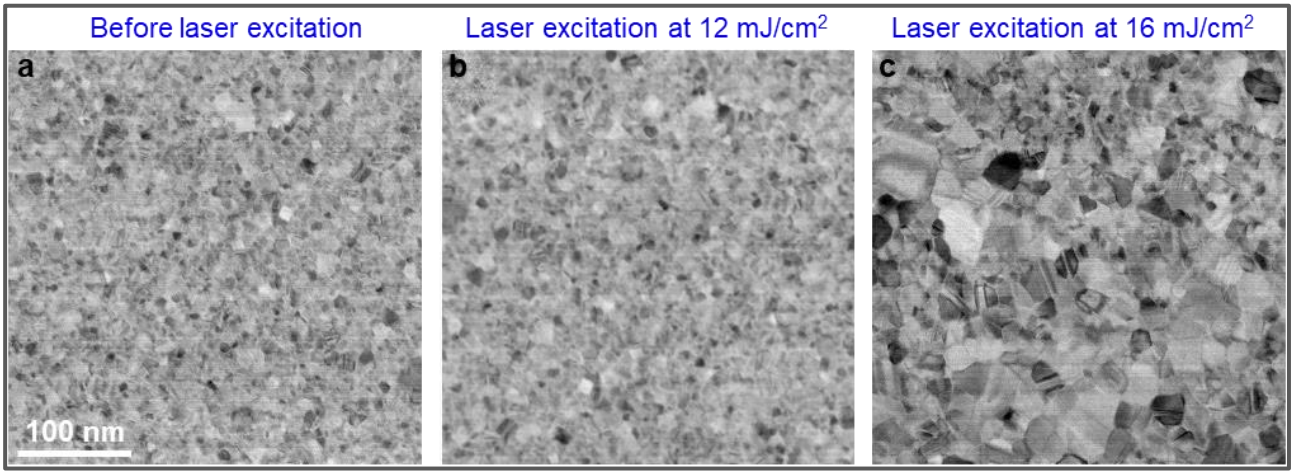

**Fig. S5. ABF images of a circular Py disk after a fs-laser pulse quenching with different fluences to show the change of the inside crystallites.** The crystallites in the Py disk show no obvious change under the excitation fluence of 12 mJ/cm$^2$ (a), while exhibit apparent growth after the fs-laser pulse excitation with the fluence of 16 mJ/cm$^2$ (c). The fs-laser pulse induced growth of crystallites induces large grain boundaries in the Py disk, which may cause more pinning sites and result in more complex magnetic structures.

**Captions for Movies S1 to S13**

**Movie S1.** Fresnel imaging of fs-laser pulse quenching induced magnetic structure change in a circular Py disk (diameter of 3 $\mu$m) at a fluence of 12 mJ/cm$^2$.

**Movie S2.** Fresnel imaging of fs-laser pulse quenching induced magnetic structure change in a square Py disk (edge length of 3 $\mu$m) at a fluence of 12 mJ/cm$^2$.

**Movie S3.** Fresnel imaging of fs-laser pulse quenching induced magnetic structure change in a triangular Py disk (edge length of 1.7 $\mu$m) at a fluence of 12 mJ/cm$^2$.

**Movie S4.** Fresnel imaging of fs-laser pulse quenching induced magnetic structure change in a circular Py disk (diameter of 1.7 $\mu$m) at a fluence of 12 mJ/cm$^2$.

**Movie S5.** Micromagnetic simulation on the magnetization relaxation dynamics of the formation of a single magnetic vortex structure in the triangular Py disk (edge length of 1.7 $\mu$m) after a fs-pulse quenching (fluence of 12 mJ/cm$^2$).

**Movie S6.** Micromagnetic simulation on the magnetization relaxation dynamics of the formation of a magnetic structure with two magnetic vortices in the triangular Py disk (edge length of 1.7 $\mu$m) after a fs-pulse quenching (fluence of 12 mJ/cm$^2$).

**Movie S7.** Micromagnetic simulation on the magnetization relaxation dynamics of the formation of a magnetic structure with three magnetic vortices in the triangular Py disk (diameter of 3.0 $\mu$m) after a fs-pulse quenching (fluence of 12 mJ/cm$^2$).

**Movie S8.** Micromagnetic simulation on the magnetization relaxation dynamics of the formation of a single magnetic vortex structure in the circular Py disk (diameter of 3.0 $\mu$m) after a fs-pulse quenching (fluence of 12 mJ/cm$^2$).

**Movie S9.** Micromagnetic simulation on the magnetization relaxation dynamics of the formation of a magnetic structure with four magnetic vortices in the circular Py disk (diameter of 3.0 $\mu$m) after a fs-pulse quenching (fluence of 12 mJ/cm$^2$).

**Movie S10.** Micromagnetic simulation on the magnetization relaxation dynamics of the formation of a magnetic structure with four magnetic vortices in the circular Py disk (diameter of 3.0 $\mu$m) after a fs-pulse quenching (fluence of 12 mJ/cm$^2$).

**Movie S11.** Micromagnetic simulation on the magnetization relaxation dynamics of the formation of a single magnetic vortex structure in the square Py disk (edge length of 3 $\mu$m) after a fs-pulse quenching (fluence of 12 mJ/cm$^2$).

**Movie S12.** Micromagnetic simulation on the magnetization relaxation dynamics of the formation of a magnetic structure with three magnetic vortices in the square Py disk (edge length of 3 $\mu$m) after a fs-pulse quenching (fluence of 12 mJ/cm$^2$).

**Movie S13.** Micromagnetic simulation on the magnetization relaxation dynamics of the formation of a magnetic structure with four magnetic vortices in the square Py disk (edge length of 3 $\mu$m) after a fs-pulse quenching (fluence of 12 mJ/cm$^2$).